\documentclass[twocolumn,showpacs,fleqn,nofootinbib]{revtex4}
\usepackage{amsmath}
\usepackage{graphicx}
\usepackage{caption}
\usepackage{subcaption}
\usepackage{float}
\usepackage{verbatim}
\usepackage{color}
\usepackage{comment}
\usepackage{hyperref}
\usepackage{epstopdf}
\usepackage{booktabs}
\usepackage[utf8]{inputenc}

\def\lsim{\mathrel{\rlap{\lower4pt\hbox{\hskip1pt$\sim$}}
    \raise1pt\hbox{$<$}}}                
\def\gsim{\mathrel{\rlap{\lower4pt\hbox{\hskip1pt$\sim$}}
    \raise1pt\hbox{$>$}}}                

\begin{document}
	
	\title{Investigating the QCD dynamical entropy in high-energy hadronic collisions}
	\author{G.S. Ramos and M.V.T. Machado}
	
	\affiliation{High Energy Physics Phenomenology Group, GFPAE  IF-UFRGS \\
		Caixa Postal 15051, CEP 91501-970, Porto Alegre, RS, Brazil}

\begin{abstract}
The dynamical entropy of dense gluonic states at in proton-proton collisions at high energies is studied by using phenomenological models for the unintegrated gluon distribution. The corresponding transverse momentum probability distributions are evaluated in terms of rapidity. The dynamical entropy density is obtained in the rapidity range relevant for the collisions at the Large Hadron Collider. The total entropy density for the dense system is computed as a function of the rapidity evolution $\Delta Y = Y-Y_0$ given a initial rapidity $Y_0$. The theoretical uncertainties are investigated and comparison with related approaches in literature is done.
\end{abstract}	

\maketitle
	
\section{Introduction}	

In high energy regime statistical physics concepts and methods are being increasingly used to describe the
outcome of particle collisions \cite{Munier:2009pc,Iancu:2004es,Le:2022exz}. It is well known that produced particle multiplicities in proton-proton and heavy ion collisions are connected with the entropy produced by these reactions \cite{Muller:2003cr,Muller:2011ra,Hanus:2019fnc,Habashy:2021qku,Matsuda:2022hok}. In this context, a current hot topic is the relation between decoherence/entanglement and  high-energy quantum chromodynamics (QCD) processes \cite{Aidala:2021pvc,Berges:2020fwq,Fu:2021jhl,Akkelin:2013jsa,Hayata:2020xxm,Baker:2017wtt,Fries:2008vp}. In general, the analyzes are complex since there are interdisciplinary connections of distinct steps of the thermalization process to non-equilibrium dynamics (see for instance the pedagogical review of Ref. \cite{Berges:2020fwq}). Another topical subject is  the entanglement entropy, which quantifies the level of entanglement between different subsets of degrees of freedom in a quantum state. Different theoretical techniques in QCD have been employed to address the  entanglement entropy of partons \cite{Kharzeev:2017qzs,Afik:2020onf,Peschanski:2016hgk,Peschanski:2019yah,Neill:2018uqw,Kutak:2011rb,Hagiwara:2017uaz,Kovner:2015hga,Armesto:2019mna,Li:2020bys,Duan:2020jkz,Duan:2021clk,Tu:2019ouv,Ramos:2020kaj,Ramos:2020kyc,Germano:2020nbh,Germano:2021brq,Feal:2020myr,Kharzeev:2021yyf,Gotsman:2020bjc,Hentschinski:2021aux,Dvali:2021ooc,H1:2020zpd,Zhang:2021hra,Liu:2022ohy,Dumitru:2022tud,Liu:2022hto}, including different interpretation for the corresponding entropy \cite{Kharzeev:2021nzh}. These rich phenomenological studies give rise to fresh developments at the confluence of quantum technologies and fundamental high energy physics \cite{Bass:2021bjv}.

In this work we focus on the \textit{dynamical entropy} for dense QCD states of matter first proposed in Ref.  \cite{Peschanski:2012cw}. Based on statistical physics tools for far-from-equilibrium processes the entropy is written as an overlap functional between the gluon distribution at different total rapidities $Y$ and saturation radius, $R_s(Y)=1/Q_s(Y)$, where $Q_s(Y)\sim e^{\lambda Y}$ is the saturation scale. In the weak coupling regime the dynamical entropy characterizes the change of the color correlation length $R_s(Y_0)\rightarrow R_s(Y)$, mirroring  the rapidity evolution $Y_0\rightarrow Y$ of a dense gluon state. The entropy functional $\Sigma^{Y_0\rightarrow Y}$ is defined in terms of the gluon transverse momenta probability distribution, $P(Y,k_{\perp})$. This distribution is defined by means of the QCD unintegrated gluon distribution (UGD), $\phi (Y,k_{\perp})$. It was shown that the total dynamical entropy density, $dS_D/dy$ is proportional to $\Sigma^{Y_0\rightarrow Y}$ and the effective gluonic degrees of freedom. A macroscopic formalism for obtaining the thermodynamic entropy associated with the production of gluons in a dilute-dense system within the Color Glass Condensate (CGC) approach was proposed in  Ref. \cite{Kutak:2011rb}. One key feature is that entropy behaves like multiplicity of produced gluons and an upper bound exists. It has been conjectured in \cite{Peschanski:2012cw} that the dynamical entropy is related to the microscopic definition of entropy based on the underlying dynamics of the CGC.  The formalism also has been extended to the initial pre-equilibrium state of a heavy ion collision. Some preliminary applications were done in \cite{Peschanski:2012cw} by using a Gaussian CGC model presenting geometric scaling property as also considered in Ref. \cite{Kutak:2011rb}. Our goal in this work is to analyse the dynamical entropy by using realistic models for the gluon UGD going beyond the Gaussian CGC approach. The rapidity dependence coming from distinct phenomenological models is investigated. Comparison with decoherence entropy is performed.

This paper is organized as follows. In next section, we  briefly review the definitions of the dynamical entropy, $\Sigma^{Y_0\rightarrow Y}$, in hadronic scattering by using the seminal work of Ref. \cite{Peschanski:2012cw}. We compute the the transverse momentum probability distributions considering realistic phenomenological models for the unintegrated gluon distribution (UGDs), $\phi(Y,k_{\perp})$, for protons. We focus  on analytical models in order to study the main features of the dynamical entropy derived from them. The corresponding  evolution in rapidity, $\Delta Y = Y-Y_0$, is investigated given an initial rapidity $Y_0$.  The total dynamical entropy density, $\frac{1}{\pi R_p^2}\frac{dS_D}{dy}$, is also introduced.  In Sec. \ref{sec3} the main results are presented  and the uncertainties associated to the formalism and possible future applications are discussed. In Sec. \ref{sec:conc}  we summarize the main results.

\begin{figure*}[t]
\centering
\begin{subfigure}{.5\textwidth}
  \centering
  \includegraphics[width=1.0\linewidth]{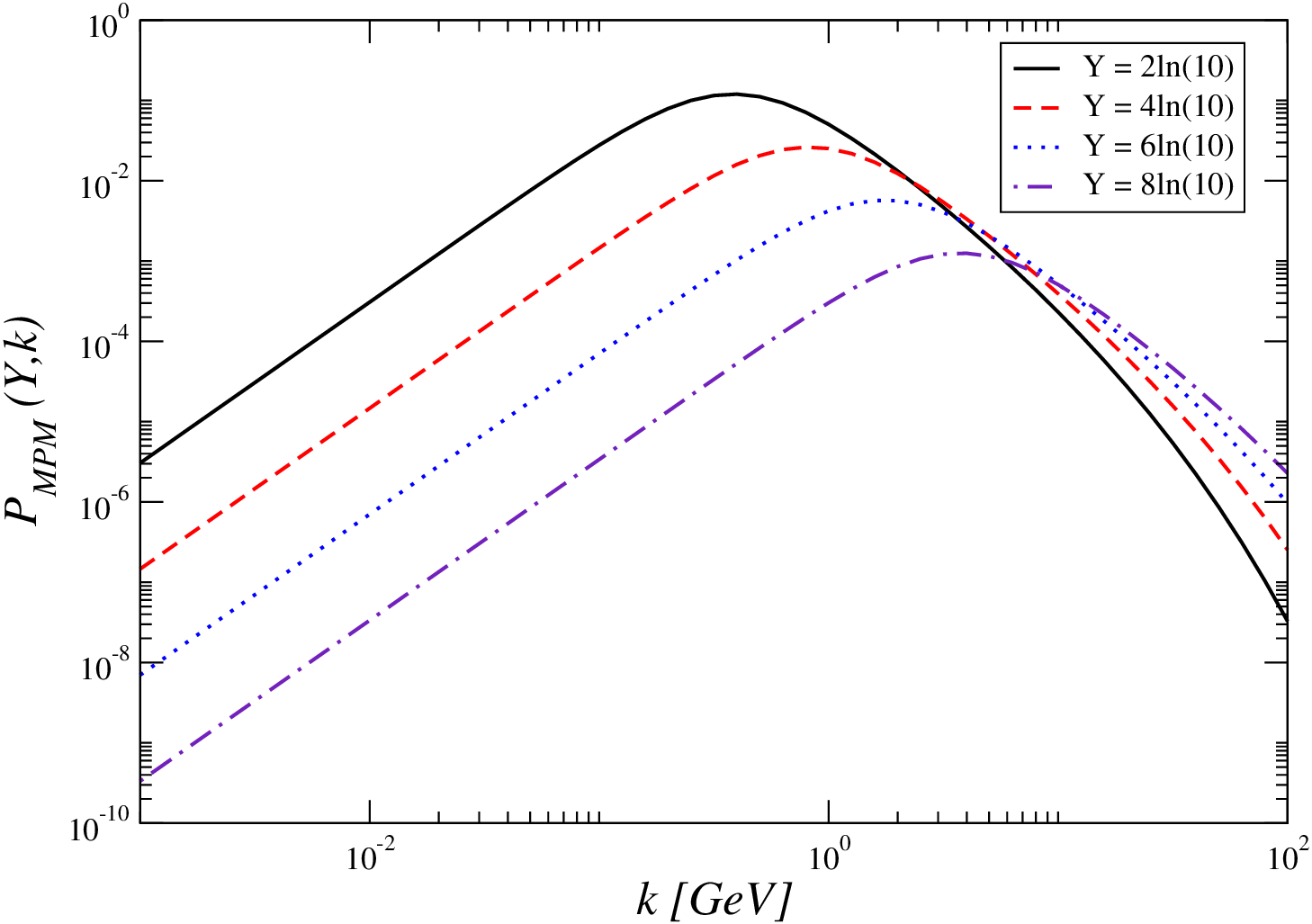}
  \caption{}
  \label{fig:sub1}
\end{subfigure}%
\begin{subfigure}{.5\textwidth}
  \centering
  \includegraphics[width=1.0\linewidth]{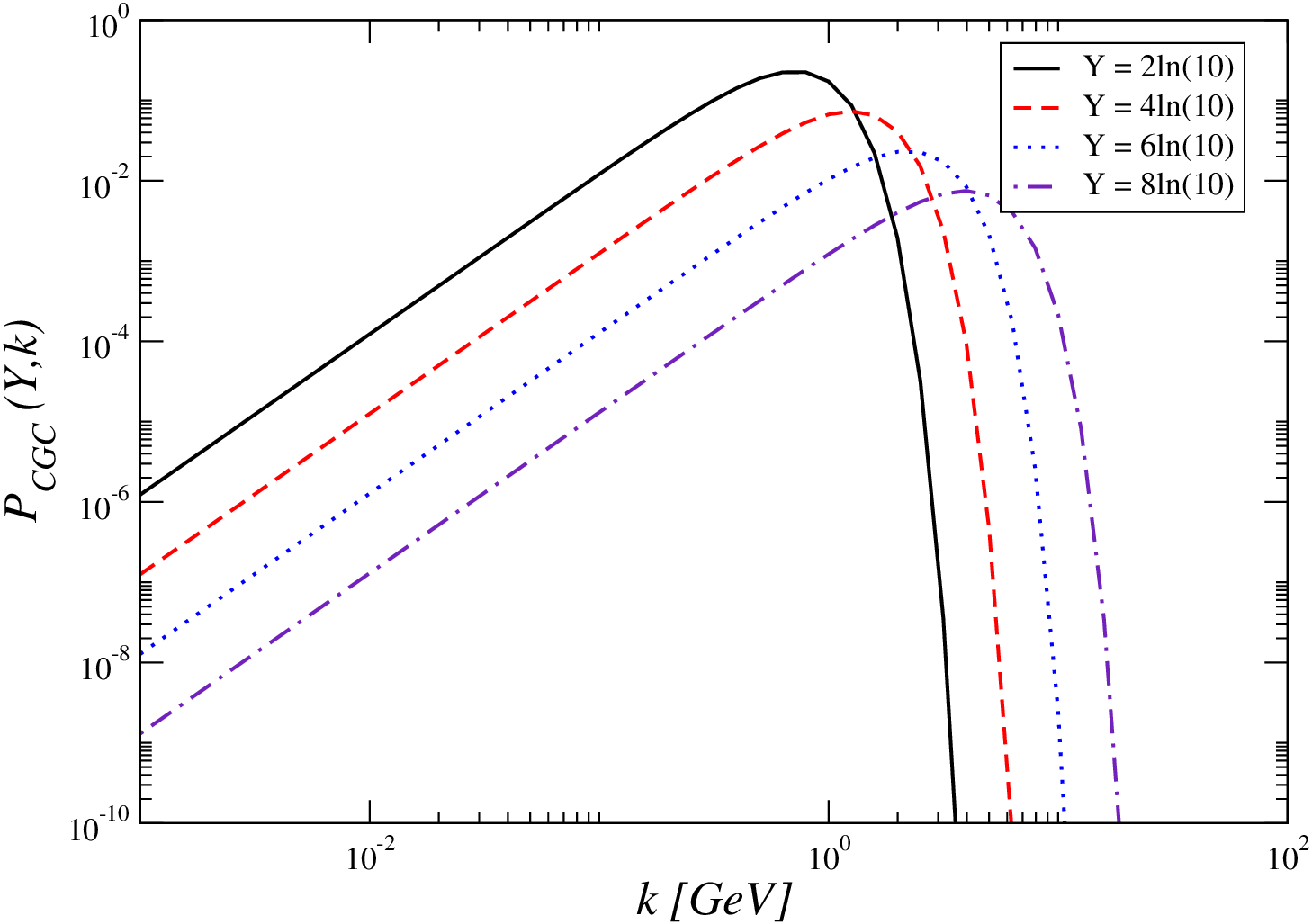}
  \caption{}
  \label{fig:sub2}
\end{subfigure}
\caption{The transverse momentum probability, $P(Y,k_{\perp})$, as a function of $k_{\perp}$ for fixed values of $Y = \ln(1/x)$ ($x=10^{-8}-10^{-2}$). Results for MPM model (left panel) and CGC Gaussian model (right panel).}
\label{fig:1}
\end{figure*}
\section{Theoretical formalism}
\label{sec2}

First, we shortly review the formulation of the dynamical entropy of dense QCD states first proposed  in Ref. \cite{Peschanski:2012cw}. Based on the classical far-from-equilibrium thermodynamics concepts in statistical physics the microscopical entropy is defined in terms of the rapidity evolution of the unintegrated parton distribution function (UGD), $\phi (Y,k_{\perp})$. Namely, the gluon dipole transverse momentum  distribution (dipole TMD) is the main input as it is associated to the suppression of the color radiation for $k_{\perp}\rightarrow 0$ and presents typically a maximum around the hadron saturation scale, $Q_s$. This last property resembles the classical notion of a gas of partons in a box of size equal to the saturation radius, $R_s = 1/Q_s$. The saturation scale can be parametrized in terms of rapidity like $Q_s(Y) = k_0^2e^{\lambda Y}$. The classical behavior here is achieved by QCD dynamics for large occupation number described by the CGC effective theory. The stating point is the definition of the transverse momentum probability distribution of the high density gluon states in the hadron target \cite{Peschanski:2012cw}:
\begin{eqnarray}
P(Y, k_{\perp}) & = &  \frac{1}{{\cal{N}}}\,\phi (Y,k_{\perp}), \label{kt-distr}\\ {\cal{N}} & = & \int  \phi (Y,k_{\perp})d^2k_{\perp},\quad
 \int  P(Y,k_{\perp})d^2k_{\perp}  =  1. \nonumber 
\end{eqnarray}
The integration over transverse momentum is simplified in case of the dipole TMD to present geometric scaling property, i.e. $\phi (Y,k_{\perp})=\phi (\tau)$, where $\tau = k_{\perp}^2/Q_s^2$ is the scaling variable. The classical statistic physics analogies considered in \cite{Peschanski:2012cw} to propose the quantities appearing in Eqs. (\ref{kt-distr}) are the probability distribution for stationary states and the Hatano-Sata identity. The transverse momenta play the role of the phase space and the rapidity $Y$ represents the dynamical parameter. The classical compression corresponds to the shrinkage of the color correlation length due to the rapidity evolution.

The dynamical entropy density of a dense system at rapidity $Y$ is defined by using the distribution $P(Y,k_{\perp})$ in  the following way \cite{Peschanski:2012cw}:
\begin{eqnarray}
\Sigma^{Y_0\rightarrow Y} &=& \left\langle \ln\left( \frac{P(Y,k_{\perp})}{P(Y_0,k_{\perp})}\right)  \right\rangle_Y \nonumber \\
&\equiv &  \int  P(Y,k_{\perp})\,\ln\left( \frac{P(Y,k_{\perp})}{P(Y_0,k_{\perp})}\right) \,d^2k_{\perp} 
\label{dyn-entropy}
\end{eqnarray}	
where $\langle \ldots \rangle_Y$ is the average over the probability distribution in the final state at $Y$. The evolution starts from initial rapidity $Y_0$ in a transverse area approximately given by the initial color correlation size, $R_0=Q_s^{-1}(Y_0)$. The dynamical entropy in Eq. (\ref{dyn-entropy}) presents the property of positivity and for distributions containing geometric scaling an analogy with the Jarzynski identity in non-equilibrium thermodynamics is possible. 

By making use of the dynamical entropy expression, the total dynamical entropy density, $dS_D/dy$, for a density state of overall transverse size $R_h$ can be computed as follows \cite{Peschanski:2012cw},
\begin{eqnarray}
\frac{1}{S_h}\frac{dS_D}{dy} = \frac{C_m}{S_0}\mu \,\Sigma^{Y_0\rightarrow Y}
 \label{total-entropy}
\end{eqnarray}
where $S_h = \pi R_h^2$ and $S_0=\pi R_0^2 = \pi Q_s^{-2}(Y_0)$. The constant $C_m = N_c^2-1/(4\pi N_c\alpha_s)$ refers to the product of the color multiplicity $(N_c^2-1)$ by the typical gluon occupation number $n_g\sim 1/4\pi N_c\alpha_s$ and $\mu$ being the effective number of partonic degrees of freedom inside a transverse cell at $Y_0$. The last quantity was identified as $\mu = 3\pi/2$ \cite{Peschanski:2012cw} by directly comparing the expressions (\ref{total-entropy}) with the macroscopic entropy proposed in Ref. \cite{Kutak:2011rb} in the dilute-dense configuration in proton-proton collisions.

In Ref. \cite{Peschanski:2012cw}, the author considered a class of Gaussian CGC models in order to estimate the dynamical entropy in a quantitative way. Namely, the $k_{\perp}$-probability distribution is supposed to present geometric scaling, $P_{\mathrm{GS}}(Y,k_{\perp})=P_{\mathrm{GS}}(\tau) \propto \Gamma^{-1}(\kappa) \tau^{\kappa -1 }e^{-\tau}$. The overlap parameter $\kappa$ parametrizes the low transverse momenta limit of the dipole TMD and in case of $\kappa =2$ the color transparency property is recovered in the gluon UGD. 

The main goal here is to compute numerically the dynamical entropy by using tested phenomenological models which describe high energy data. Specially, those models with a accurate description of the low transverse momentum behavior of the gluon distribution. We prefer analytical models in order to single out the main features of the total dynamical entropy density. A phenomenological model that accounts for the geometric scaling present in charged hadrons production in $pp$ collisions combined with a Tsallis-like distribution observed from the hadron spectrum measured is proposed in Refs.  \cite{Moriggi:2020zbv,Moriggi:2020qla} (hereafter MPM model). The corresponding gluon UGD is expressed as:
\begin{eqnarray}
\phi_{\mathrm{MPM}}(Y,k_{\perp})=\frac{3\,\sigma_{0}}{4\pi^2\alpha_{s}}\frac{\tau\,\beta (\tau )}{\left(1+\tau\right)^{1+\beta(\tau)}},
\label{FMPM}
\end{eqnarray}
In expression above, $\alpha_{s} = 0.2$, $Q_{s}^2(Y) = k_0^2e^{0.33 Y}$  with $k_0^2= \bar{x}_0^{0.33}$ GeV$^2$. The power-like behavior of the gluons produced at high momentum spectrum is determined via the function $\beta (\tau) = a\tau^{b}$ where $\tau$ is the scaling variable. The set of parameters, $\sigma_{0}$, $\hat{x}_{0}$, $a$ and $b$ are fitted from DIS data available at small-$x$. In calculations we consider the parameters from Fit B in Ref. \cite{Moriggi:2020zbv}. Namely, $\sigma_{0} = 20.47$ mb, $\bar{x}_{0} = 3.52 \times 10^{-5}$, $a = 0.055$ and $b = 0.204$. By using the definition for the transverse momentum probability distribution and the manifest geometric scaling property contained in the MPM model one obtains,
\begin{eqnarray}
P_{\mathrm{MPM}}(Y,k_{\perp}) = \frac{1}{\pi Q_s^2(Y)\xi}\frac{\tau\,\beta (\tau )}{\left(1+\tau\right)^{1+\beta(\tau)}},
\label{PMPM}
\end{eqnarray}
where $\xi = 4.34618$ results from the numerical calculation of the normalization ${\cal{N}}$ in Eq. (\ref{kt-distr}). 

We will compared the MPM model with the one proposed in Ref. \cite{Kutak:2011rb}, which is the baseline model for the investigation in Ref. \cite{Peschanski:2012cw} (hereafter CGC Gaussian model). There the saturation scale is associated to the maximum of the gluon density and the cross section for the inclusive gluon production is computed in dilute-dense regime in proton-proton collisions. The saturation scale drives the entropy of produced gluons,  which is written in terms of multiplicity of these partons. In such a macroscopic definition of entropy, it behaves like number of partons whose distribution is given by the gluon UGD. The dipole TMD and the corresponding transverse momentum probability distribution are given by,
\begin{eqnarray}
\phi_{\mathrm{Gaus}}(Y,k_{\perp}) &= &  \frac{C_FA_{\perp}}{4\pi^2\alpha_s}\tau e^{-\tau/2},\\
P_{\mathrm{Gaus}}(Y,k_{\perp})  & = &  \frac{\tau e^{-\tau/2}}{4\pi Q_s^2(Y)},
\label{PCGCGAUS}
\end{eqnarray}
with $A_{\perp} = \pi R_p^2$ being the proton transverse area. For the saturation scale scale the following parametrization has been used, $Q_s^2(Y) = k_0^2e^{\lambda Y}$, with $k_0^2= \bar{x}_0^{\lambda}$ GeV$^2$  ($\bar{x}_0=4 \times 10^{-5}$ and $\lambda = 0.248$) \cite{Golec-Biernat:2017lfv}. The expression above corresponds to an overlap parameter $\kappa =2$ for $P_{\mathrm{GS}}(\tau )$ as discussed before. Despite being very simple, this GBW-like parametrization is not able to describe the charged hadrons $p_T$-spectra due to the highly suppressed exponential tail at large $k_{\perp}$. This characteristic feature has been demonstrated in Ref. \cite{Szczurek:2003fu}. 

In order to analyse a dipole TMD containing more physical information and having the correct theoretical behavior for small and large transverse momenta we will investigate the one obtained in Refs. \cite{Abir:2017mks,Siddiqah:2018qey}. It was derived as a general form of solution of $\phi(Y,k_{\perp})$ which reproduces both McLerran-Venugopalan initial conditions and Levin-Tuchin solution in their appropriate limits. It connects both limits smoothly and better approximates the numerical solution of full leading order Balitsky-Kovchegov equation,  specially deep in the saturation region. In this limit the dipole gluon TMD goes
to zero as $k_{\perp}\rightarrow 0$. The results present similarity with the Sudakov form factor \cite{Siddiqah:2018qey}. Initiating with dipole TMD at small transverse momentum from Levin-Tuchin (LT) solution of S-matrix, the corresponding UGD takes the form in the region $Q_s\gsim k_{\perp}\gsim \Lambda_{QCD}$ \cite{Siddiqah:2018qey}:
\begin{eqnarray}
\phi_{\mathrm{LT}}^{\mathrm{sat}}(Y,k_{\perp}) = - \frac{N_cA_{\perp}\varepsilon}{\pi^3 \alpha_s} \ln \left(\frac{\tau}{4}  \right)\exp \left[-\varepsilon \ln^2 \left(\frac{\tau}{4}  \right)   \right],
\label{Levin-Tuchin}
\end{eqnarray}
where $\phi_{\mathrm{LT}}^{\mathrm{sat}}$ has been obtained at small transverse momentum in terms of a series of Bells polynomials. The expression above corresponds to the leading logarithmic approximation for the resummed series and the constant $\varepsilon \approx 0.2$ arises from
the saddle point condition along the saturation border.  Outside of the saturation boundary, $k_{\perp}\gsim Q_s$, but close to the saturation line, the QCD color dipole
amplitude in transverse size space has the form $N(r,Y)\approx (r^2Q_s^2)^{\gamma_{s}}$ ($\gamma_s \approx 0.63$ is the value of the effective anomalous dimension in the vicinity of the saturation line). In this limit, the dipole TMD can be written as:
\begin{eqnarray}
\phi_{dip}^{\mathrm{dil}}(Y,k_{\perp}) \propto \frac{N_cA_{\perp}\varepsilon}{\pi^3 \alpha_s} \tau^{-\gamma_s}.
\label{ugd-dilute}
\end{eqnarray}
For large transverse momenta violation of the geometric scaling is expected and the typical BFKL diffusion term appears (see, for instance the phenomenology associated to the AGBS model \cite{Amaral:2020xqv} and references therein).

Based on the theoretical dipole TMD features summarized by Eqs. (\ref{Levin-Tuchin})-(\ref{ugd-dilute}), we propose the following expression for the transverse momentum probability distribution,
\begin{eqnarray}
 P_{\mathrm{LT}} (Y,k_{\perp})= \left \{ \begin{matrix} -B \ln \left(\frac{\tau}{4}  \right)\exp \left[-\varepsilon \ln^2 \left(\frac{\tau}{4}  \right)   \right], & \tau<1, \\B (d\tau)^{-\gamma_s}\exp \left[-\varepsilon \ln^2 \left(\frac{\tau}{4}  \right)   \right], & \tau \geq 1, \end{matrix} \right. \nonumber\\
 \label{PLT}
\end{eqnarray}
where $d = [\ln(4)]^{-\frac{1}{\gamma_s}}$ and $B\simeq 0.1/\pi Q_s^2(Y)$ is the overall normalization. In the region $\tau >> 1$, the suppression factor in second line of (\ref{PLT}) has a twofold goal: it ensures the convergence of $k_{\perp}$-integration in Eq. (\ref{kt-distr}) and helps on the function continuity at $\tau=1$. Fortunately, the large-$k_{\perp}$-tail has no severe effect in the calculation of the dynamical entropy since the integrand is dominated by the contribution around $\tau=1$. The same parametrization for the saturation scale as used in CGC-Gaussian model has been considered.

In next section we compute numerically the dynamical entropy for the models discussed above as well as the corresponding total dynamical entropy density. The rapidity range investigated is the relevant one for proton-proton collisions at the LHC.

\section{Results and discussions}
\label{sec3}
\begin{figure}[t]
\includegraphics[scale=0.35]{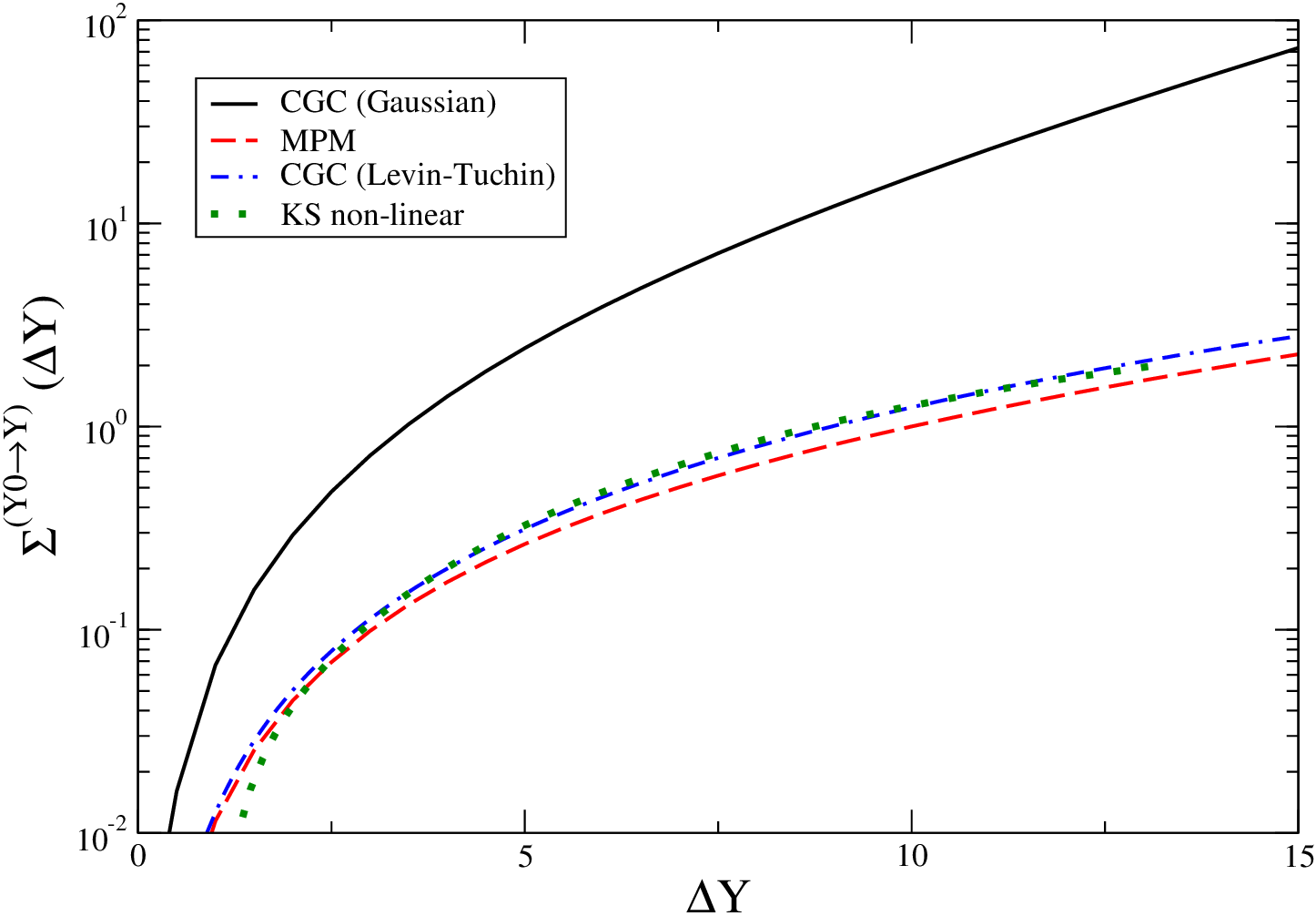}
\caption{Dynamical entropy corresponding to the QCD evolution in rapidity  $Y_0\rightarrow Y$, with $\Delta Y = Y-Y_0$. Initial rapidity has been set at $Y_0=\ln(1/x_0)$, where $x_0=10^{-2}$. Numerical results for MPM (dashed line), CGC Gaussian (solid line) and CGC Levin-Tuchin (dot dashed line) models are shown.}
\label{fig:2}
\end{figure}

In Fig. \ref{fig:1} is shown the transverse momentum probability distribution of gluons for the MPM (\ref{fig:1}-a) and CGC Gaussian (\ref{fig:1}-b) UGD models. It is presented the $k_{\perp}$-dependence for several rapidity values, $Y/\ln(10)=2,4,6,8$, which correspond to longitudinal momentum fraction for gluons, $x=10^{-8}-10^{-2}$. Both models present geometric scaling property, $\phi (Y,k_{\perp})=\phi (\tau = k_{\perp}^2/Q_s^2(Y))$, and the peak occurs at transverse momentum proportional to the saturation scale.However, in each models its location is different: in the CGC Gaussian $k_{\perp}^{max}=\sqrt{2}Q_s(Y)$ whereas for MPM it corresponds to $k_{\perp}^{max}\approx \sqrt{0.954}Q_s(Y)$.  We think these model samples are enough for a qualitative and quantitative analyzes. It is clear that for models based on scaling the main contribution to statistical averages comes from the integration region around saturation scale. The behavior for $k_{\perp }> Q_s(Y)$ is distinct as the CGC Gaussian model has a strong exponential suppression on $k_{\perp}^2$, which is not the case for the MPM model. These features will be the origin for a diverse behavior of the dynamical entropy coming from the models.

Now the dynamical entropy of partons (gluons) at rapidity $Y$ obtained from the QCD evolution $Y_0\rightarrow Y$ is computed. We consider the initial rapidity at $Y_0=\ln(1/x_0)$ with $x_0=10^{-2}$. The values of $x\leq x_0$ corresponds to the limit of validity for the application of the phenomenological UGDs models considered here. At this initial rapidity partons populate a transverse area proportional to the initial color correlation size $R_0(Y_0)=1/Q_s(Y_0)$. In Fig. \ref{fig:2} the dynamical entropy, $\Sigma^{Y_0\rightarrow Y}$, Eq. (\ref{dyn-entropy}), is presented as a function of the rapidity difference $\Delta Y = Y-Y_0$. The results are presented for the analytical  MPM model (dashed line), the CGC Gaussian model (solid line) and the CGC Levin-Tuchin model (dot-dashed line). Numerical solutions for the UGD coming from non-linear evolution equations are subsequently discussed.   It is verified the MPM and CGC Levin-Tuchin models are almost coincident meaning that the phenomenological MPM UGD mimics correctly the expected theoretical behavior of the dipole UGD in the saturation region. The CGC Gaussian model provides a steeper growth of the dynamical entropy in terms of rapidity $Y$ compared to other models. In order to check out the salient differences, the entropy for the Gaussian model is given by:
\begin{eqnarray}
\Sigma_{\mathrm{Gaus}}^{Y_0\rightarrow Y} &=& 2\left[\left(\frac{Q_s^2(Y)}{Q_s^2(Y_0)}-1 \right) -\ln \left(\frac{Q_s^2(Y)}{Q_s^2(Y_0)} \right)    \right],\\
& = & 2\left( e^{\lambda \Delta Y} -1 - \lambda \Delta Y \right),
\end{eqnarray}
where in our case $Q_s^2(Y)= Q_s^2(Y_0)e^{\lambda \Delta Y }$. From the expression above, the asymptotic dependence at large rapidities is $\Sigma_{\mathrm{Gauss}}\simeq e^{\lambda \Delta Y}$ (with a linear function of $\Delta Y$ in a logarithmic scale, $\ln (\Sigma_{Gauss})\propto 0.3 \Delta Y$). It is seen that the large $Y$ limit is practically independent of the initial conditions, $Y_0$ and $Q_s(Y_0)$. On the other hand, the MPM model gives rise to an entropy parametrized as follows:
\begin{eqnarray}
\Sigma_{\mathrm{MPM}}^{Y_0\rightarrow Y} \approx  (1+\gamma_s)\left( e^{\delta \Delta Y} - g - \delta \Delta Y \right),
\end{eqnarray}
where $\delta \simeq 0.088$ and $g\simeq 0.95$ in the range $\Delta Y>>5$. 

The total dynamical entropy density $dS_D(Y_0\rightarrow Y)/dy$ is also computed. It is explicitly obtained from the dynamical entropy $\Sigma^{Y_0\rightarrow Y}$ \cite{Peschanski:2012cw}:
\begin{eqnarray}
\frac{dS_D}{dy} = \frac{3}{8}\frac{N_c^2-1}{N_c\alpha_s}\frac{S_p}{S_0}\,\Sigma^{Y_0\rightarrow Y} ,
\end{eqnarray}
where $S_p = \pi R_p^2$ is the transverse hadronic target size and $S_0=\pi R_0^2$. The initial transverse cell at rapidity $Y_0$ is denoted by $S_0$, with $R_0 = Q_s^{-1}(Y_0) $ being the typical correlation length for the proton at rest. The overall normalization condenses all information on the degrees of freedom  as the color multiplicity, the parton occupation number in longitudinal coordinate space and the average number of parton degree of freedom inside a transverse cell. The expression above does not include parton correlation effects \cite{Peschanski:2012cw}.  For the CGC Gaussian model the total dynamical entropy is given by:
\begin{eqnarray}
\frac{dS_D^{\mathrm{Gaus}}}{dy} & =& \frac{3}{2}\frac{(N_c^2-1)}{N_c\alpha_s}\frac{S_p}{S_0} \left[\frac{Q_s^2(Y)}{Q_s^2(Y_0)}-\left(1+\ln \frac{Q_s^2(Y)}{Q_s^2(Y_0)}  \right)\right], \nonumber \\
&\propto & e^{\lambda \Delta Y}\left[1+e^{-\lambda \Delta Y}\left( 1+\lambda \Delta Y \right)  \right],
\end{eqnarray}
where the main features resembles those for the dynamical entropy $\Sigma^{Y_0\rightarrow Y}$ discussed before. 

\begin{figure}[t]
\includegraphics[scale=0.35]{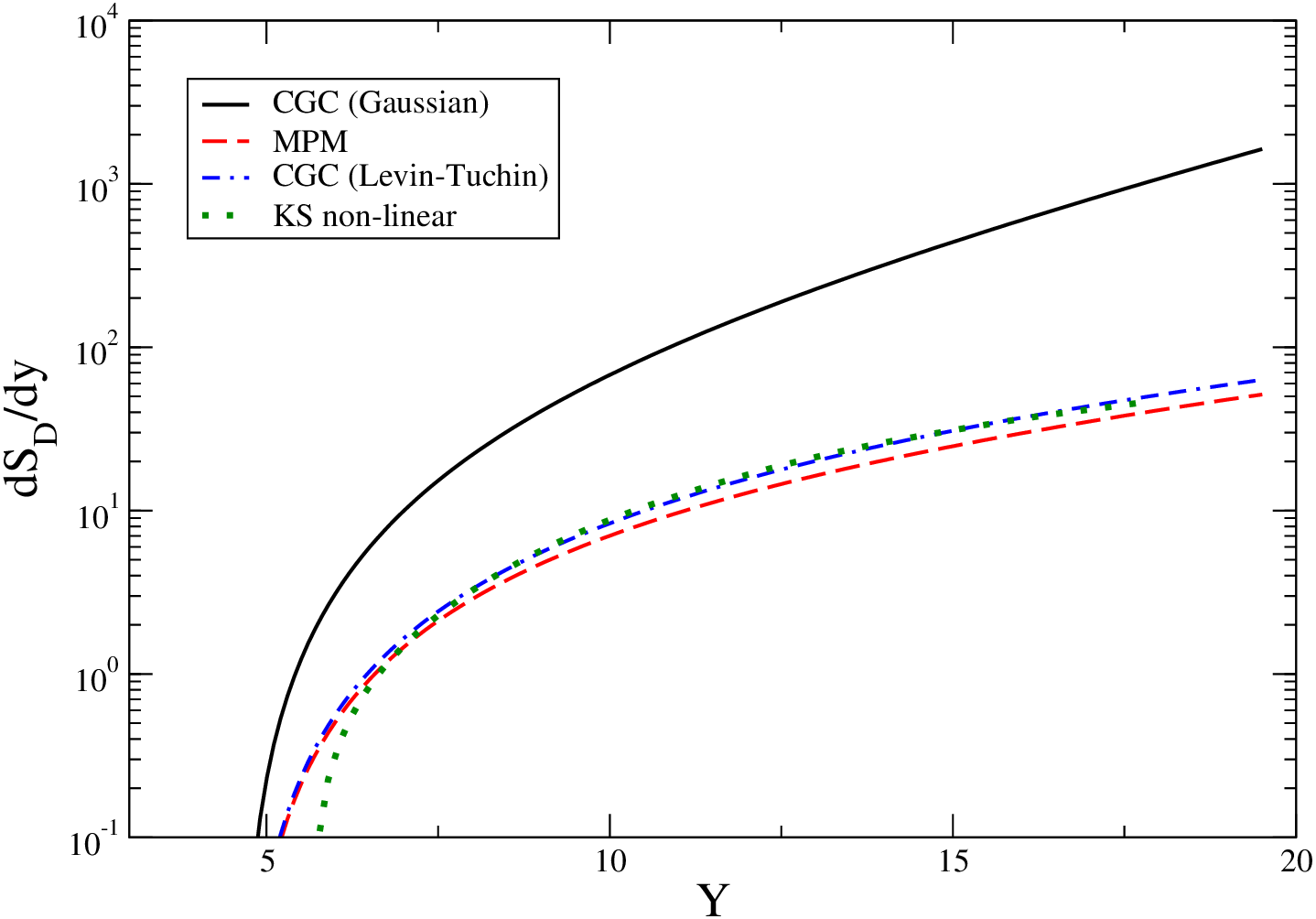}
\caption{Total dynamical entropy in proton-proton collisions corresponding to the QCD evolution in rapidity, $Y_0\rightarrow Y$, within the range $\Delta Y=[0,15]$. Same notation as previous figure.}
\label{fig:3}
\end{figure}

In Fig. \ref{fig:3} the total dynamical entropy is presented for the 3 analytical models considered here. The notation for the lines is the same as in previous figure. It is shown as a function of $Y$ in the range $\Delta Y =[0,15]$, which corresponds to a QCD evolution from $x=10^{-2}$ to $x=10^{-8}$. We have used $\alpha_s =0.2$ and $R_p = 0.8414$ fm in the numerical calculations. These values can be compared to the extracted (thermodynamic) entropy per unity of rapidity $dS/dy$ in proton-proton collisions using the Pal-Pratt method as shown in Ref. \cite{Hanus:2019fnc}. The entropy for pions in minimum bias collisions at 7 TeV is $(dS/dy)^{\pi}_{y=0}\simeq 20$ and $(dS/dy)^{\pi}_{y=0}\simeq 71$ for high multiplicity collisions. For the MPM and CGC Levin-Tuchin models, at very small-$x$ the order of magnitude is similar. For instance, in the range $\Delta Y=[10,20]$ one finds $\langle dS_D/dy \rangle \approx 40$.

In order to verify the theoretical uncertainty associated to the phenomenological models the Kutak-Sapeta model for the UGD has been considered which is  the numerical solution of the unified BK/DGLAP equation \cite{Kutak:2012rf,Kutak:2003bd} (hereafter labeled as KS nonlinear). It is an example of UGD that does not present explicit geometric scaling property, specially at large-$k_{\perp}$. The results for this UGD are presented in Figs. \ref{fig:2} and \ref{fig:3} (dotted curves). Interestingly, for large enough $\Delta Y$ the numerical results using KS non-linear are nicely mimicked by the simple parametrization CGC Levin-Tuchin. Similar exercise can be done using other numerical results for the proton UGD available in  the updated  TMDlib2 library \cite{Abdulov:2021ivr}.

As a final discussion, we would like to compare the concept of dynamical entropy with the decoherence entropy \cite{Muller:2003cr,Muller:2011ra,Fries:2008vp}.  The main idea is that dense states in nucleon or nuclei can be represented by coherent states, $|\alpha\rangle$, at the initial stages of collisions. A coherent state $|\alpha \rangle$  is defined to be the eigenstate of the annihilation operator $\hat{a}$ with corresponding eigenvalue $\alpha$. They can be expressed as a superposition of particle number eigenstates, $|n\rangle$. The large number of occupation allows to describe the corresponding fields as classical one. For a single mode of the field, the density matrix is non-diagonal $\hat{\rho}_{mn}= \langle m|\alpha \rangle \langle \alpha|n \rangle$ and $|\alpha\rangle $ has zero entropy. For this pure quantum state, $S=-\mathrm{Tr}(\hat{\rho})\ln (\hat{\rho})=0$. Afterwards, the complete decoherence  is associated  to the decay of all off-diagonal matrix elements of $\hat{\rho}$ such that $\hat{\rho}_{mn}^{\mathrm{dec}}=|\langle n|\alpha \rangle|^2$. The particle number now follows the Poisson distribution where the average number of particles is $\langle n\rangle = |\alpha|^2$. The entropy amount $S_{dec}$ of this mixed state is determined as follows \cite{Muller:2003cr,Fries:2008vp},
\begin{eqnarray}
S_{dec} & = &  \sum_{n=0}^{\infty} e^{-\langle n \rangle} \frac{\langle n \rangle^n}{n!}\ln \left(e^{-\langle n \rangle} \frac{\langle n \rangle^n}{n!}  \right),\\
& \approx & \frac{1}{2}\left[\ln \left(2\pi \langle n \rangle\right) +1-\frac{1}{6\langle n \rangle}+\ldots \right].
\label{sdec}
\end{eqnarray}

The $S_{dec}$ can be contrasted with the equilibrium entropy at temperature $T$, $S_{eq}$, for a single quantum oscillator with same average total energy. Given that the average occupation number is in this case $\langle n \rangle = (e^{\omega T}-1)^{-1}$ one obtains:
\begin{eqnarray}
S_{eq} = \ln \left(\langle n \rangle +1\right) + \langle n \rangle\ln \left(\frac{\langle n \rangle+1}{\langle n \rangle} \right),
\label{seq}
\end{eqnarray}
where $S_{eq}\approx 2S_{dec}$ for very large occupation number and close to unity for moderate $\langle n \rangle$ \cite{Muller:2003cr,Fries:2008vp}.

The field is a system of coupled oscillators and after total decoherence it can be described as an assemblage of $N$ particles. These particles have been generated by the decoherence of $N_{\alpha}$ coherent
quantum states.  Every coherent state contributes on average for $\langle n\rangle = N/N_{\alpha}$ gluons. After complete equilibration,  the decoherence entropy reaches to $S_{dec} \approx (N_{\alpha}/2)[\ln(2\pi\langle n \rangle )+1]$. In Refs. \cite{Muller:2003cr,Fries:2008vp} the  decoherence entropy was quantified for nucleus-nucleus collisions. The initial number of coherent domains per transverse area was set at $N_{\alpha}\sim Q_s^2R_A^2$ and the average number of decohering gluons per coherence domain is given by $\langle n \rangle \approx C_F\ln (2)/\pi \alpha_s^2$. Still, the longitudinal coherence length
is taken as $\Delta y \approx (\alpha_s)^{-1}$.

\begin{figure}[t]
\includegraphics[scale=0.35]{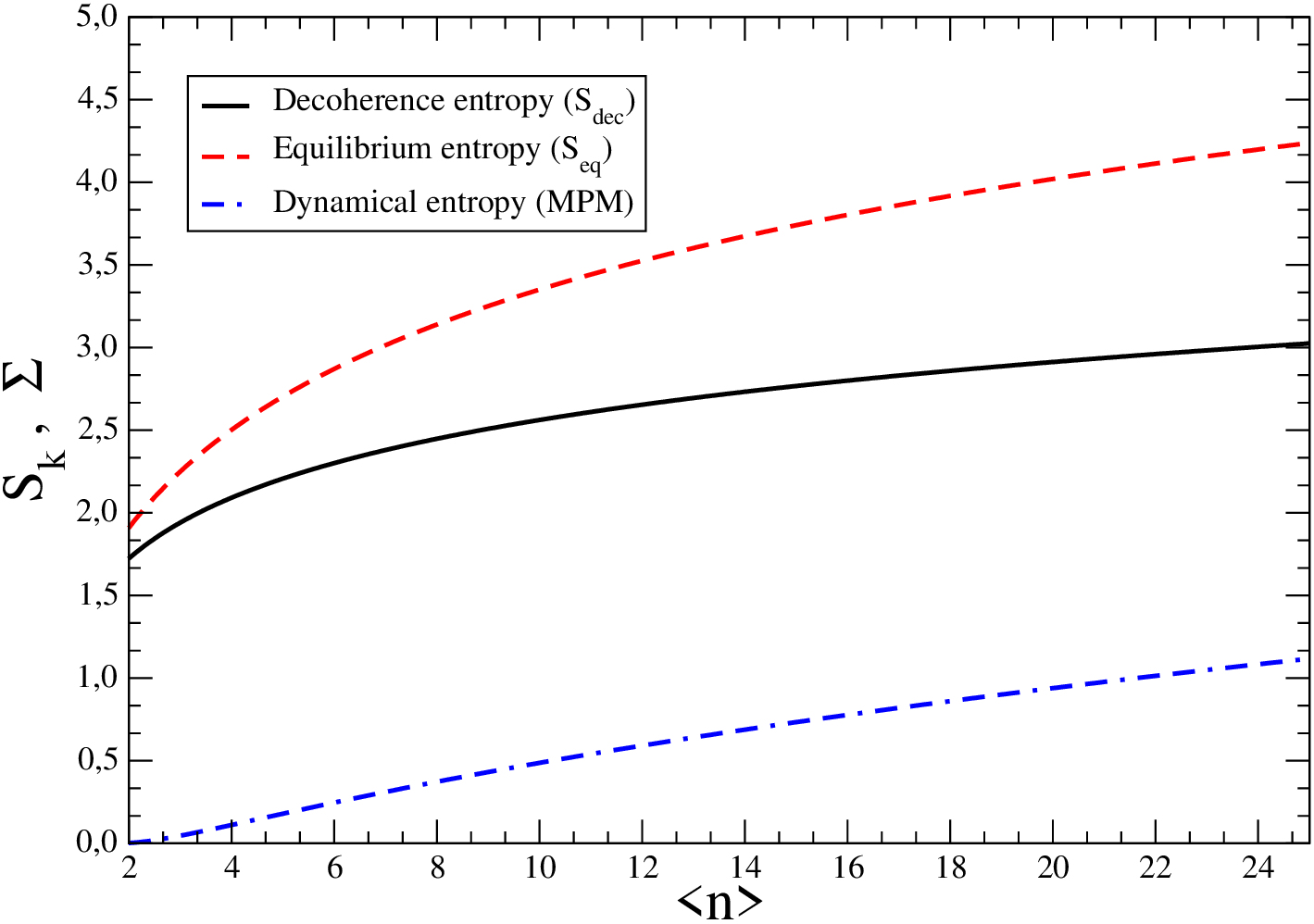}
\caption{The dynamical entropy,  $\Sigma$, as a function of average number of occupation (dotted-dashed line). Comparison with the decoherence entropy (solid line) and equilibrium entropy (dashed line) for coherent state of a single field mode os shown.}
\label{fig:4}
\end{figure}

On the other hand, it was shown in Ref. \cite{Peschanski:2012cw} that the dynamical entropy can be viewed (by using the Jarzynski identity) as the entropy production $\Delta S$ by parton degree of freedom associated to a compression $R_s(Y_0)\rightarrow R_s (Y)$ as the system relaxes to a state within the domain size $R_s(Y)$. In summary, the dynamical entropy is acquired by a gluon dense initial state through the rapidity evolution and delivered by the relaxation. In Fig. \ref{fig:4} the dynamical entropy per average number of gluonic degrees of freedom (with $Y_0$ fixed as before), $\Sigma$, is shown as a function of average number of occupation $\langle n\rangle$. Following Ref. \cite{Kharzeev:2017qzs} we considered the gluon average number given by the gluon density at scale $Q^2$, i.e. $\langle n \rangle = xG(x,Q^2)$. The resolution scale $Q^2=Q_s^2(Y)$ will be chosen for the typical transverse momentum inside hadron. The gluon density at this scale is given by $xG(x=e^Y,Q_s^2)=CQ_s^2(Y)$, with $C= 3S_p(1-2/e)/4\pi^2\alpha_s$ (see discussion about this expression in Ref. \cite{Ramos:2020kaj}). The result is shown for the MPM model. A comparison is done  with the decoherence entropy $S_{dec}$, Eq. (\ref{sdec}),  and equilibrium entropy $S_{eq}$, Eq. \ref{seq}, for coherent state of a single field mode. For large number of occupation the behavior of the equilibrium entropy of a single mode as a function of $\langle n\rangle$ is quite similar to the dynamical entropy. Of course, we noticed that a different definition for $\langle n \rangle$ had been used in  each case.

The analysis presented here can be extended to a dense system at initial state in ultra-relativistic heavy ion collisions. The main input will the gluon unintegrated distribution associated to the glasma phase. In Ref. \cite{Peschanski:2012cw} it was assumed a CGC Gaussian UGD based on geometric scaling arguments in order to approximately determine the dynamical entropy. In leading order the entropy has the form $\Sigma^{Y_0\rightarrow Y} \sim \kappa_g [Q_s^2(Y)/Q_s^2(Y_0)]$, where $\kappa_g$ is the overlap parameter incase of glasma state. This dynamical entropy can be viewed as the initial entropy density, $s(0)$ and it has connection with thermalization process in heavy ion reactions. It would be worth  consider physical parametrizations for the glasma UGD, which it will be considered in a future analysis. 

\section{Summary}
\label{sec:conc}

We have studied the QCD dynamical entropy for high energy proton-proton  collisions in the LHC energy regime, which is theoretically obtained by the the gluon dipole TMD as first proposed in Ref.\cite{Peschanski:2012cw}. The $k_{\perp}$ probability distribution has been determined by using analytical models for the gluon UGD. Namely, we consider the MPM phenomenological model  which describes accurately the charged particle spectra measured at the LHC. It presents geometric scaling and mimics the correct behavior at large (pQCD) and small (parton saturation) gluon transverse momentum. We compared the results from MPM model to those coming from the CGC framework. A gaussian model (CGC Gaussian) for the gluon UGD and the one based on the Levin-Tuchin law at low-$k_{\perp}$ (CGC Levin-Tuchin) have been investigated. Both also present geometric scaling property. In all cases the maximum of the probability distribution is located at $k_{\perp}\sim Q_s$. The corresponding dynamical entropy, $\Sigma^{Y_0\rightarrow Y}(\Delta Y)$, is computed and the total entropy density, $S_p^{-1}dS_D/dy$, as well. It is found a strong dependence on $\Delta Y$ for the CGC Gaussian model. The results for the analytical MPM and CGC Levin-Tuchin models are practically coincident. Moreover, the dynamical entropy is evaluated by using the Kutak-Sapeta model including non-linear corrections to the QCD evolution equation (KS non-linear) \cite{Kutak:2012rf}. The results for KS models are coincident with the ones obtained from the analytical parametrization CGC Levin-Tuchin.  A direct comparison of the dynamical entropy as a function of average gluon number, $\langle n \rangle$, with the decoherence entropy $S_{dec}$ \cite{Muller:2003cr,Fries:2008vp}  was performed.  Interestingly, at large $\langle n \rangle$ the behavior of the equilibrium entropy $S_{eq}$ of a single mode is similar to the dynamical entropy.

In summary, the present work the QCD dynamical entropy in hadron scattering processes is carefully studied using realistic models for the dipole TMD function. The analysis has been  helpful to single out the main aspects of the rapidity dependence of the total dynamical entropy density in proton-proton collisions. This is the stating point to investigate the dynamical entropy in the initial states of the heavy ions collisions.

\begin{acknowledgments}
 This work was  partially financed by the Brazilian funding agencies CNPq and CAPES.
\end{acknowledgments}

\end{document}